# On ad hoc routing with guaranteed delivery

Mark Braverman*

**Abstract**

We point out a simple poly-time log-space routing algorithm in ad hoc networks with guaranteed delivery using universal exploration sequences.

# 1 Introduction

## 1.1 Routing model

We consider the problem of routing in an ad hoc network. The network is comprised of independent agents each of which has very limited resources. Moreover, the topology of the network is not known a-priori to neither the agents nor the algorithm. We assume that the nodes have unique universal names (e.g. physical locations) that are drawn from a namespace of size $n$ (for example, for IPv4 addresses $n = 2^{32}$). The goal is to route a message from a source node $s$ to a target node $t$ (routing), or from $s$ to its connected component (broadcasting). The goal is to have a delivery algorithm that has high probability of success or, if possible, *guarantees* delivery. In the present paper we assume that the network is static. That is, the graph does not change during the delivery process.

Nodes do not have sufficient resources to "learn" the network. In particular, nodes are restricted to having $O(\log n)$ memory. Thus the nodes can store and operate with a finite number of indexes and identifiers, but have insufficient memory to store global information about the network, such as the list of nodes and connections. An overhead of $O(\log n)$ is allowed on top of the messages to facilitate delivery. Without the overhead, the algorithm becomes *local* and lower bounds showing that guaranteed routing is impossible in general exist [2].

Algorithms for guaranteed routing and broadcasting on restricted classes of networks, such as planar networks has been devised [1, 2, 4, 5, 9]. However, giving good algorithms with guaranteed delivery in general 3-dimensional graphs appears to be hard. In the present note we observe that using state-of-the-art tools from logspace derandomization [8], at least theoretically, the problem is solvable in polynomial time. The algorithm runs in time polynomial in the size of the connected component of $s$, which could be considerably less than $n$, which is the global bound on the number of nodes in the network. In particular, the algorithm does not require to know $n$ ahead of time.

---

*Dept. of Computer Science, University of Toronto

## 1.2 Results

The main statement is that there is a poly-time ad hoc routing algorithm with guaranteed delivery that does not require intermediate nodes to store any information, and limits the memory usage and the overhead on the messages by $O(\log n)$. Moreover, the running time is polynomial in the size of the connected component of the source node, and there is no need to know this size in advance.

**Theorem 1** *There is an ad hoc routing algorithm (on a static network) in which the nodes are allowed to use space $O(\log n)$, and the overhead on the message is also $O(\log n)$ where $n$ is the size of the namespace for the nodes. Moreover, the routing runs in time $poly(|\mathcal{C}_s|)$, where $\mathcal{C}_s$ is the connected component of the source node in the network. There is no need to know $|\mathcal{C}_s|$ in advance.*

The same algorithm works for the *broadcasting* problem, where $s$ wants to send the message to all the vertexes in its connected component.

Since the nodes are not allowed to "remember" anything about the path the algorithm tries to take, a natural, if wasteful, approach would be to try and route the message randomly from $s$ to $t$. The random routing should stop when the message reaches $t$, or alternatively, when a confirmation from $t$ reaches $s$. There are several problems with this simple approach. First of all, if we are sufficiently unlucky, the message will never reach $t$. Second, unless we are willing to deposit a token in each node the message visits along the path, there is no reliable way of returning a confirmation from $t$ to $s$. Third, if there is no path from $s$ to $t$, then the algorithm will never terminate. These problems are addressed using *universal exploration sequences*. These sequences can be seen as a derandomized version of the randomized walk. It prescribes a series of polynomially many steps that, if followed, are *guaranteed* to visit every vertex in the network. Moreover, Reingold [8] has shown that such sequences can be constructed deterministically in logarithmic space. Thus, a routing algorithm as in Theorem 1 can be obtained.

The degree of the polynomial is not optimal. It is, in fact, fairly high if one uses the construction from [8]. However, the existence of the algorithm mean that theoretically, guaranteed poly-time log-space ad hoc routing is possible irrespective of the graph's topology. Thus, it is unlikely that the problem would have good theoretical lower bounds even for very bad topologies. In practice this means that any probabilistic routing algorithm with high success probability and small expected running time, can be converted into a routing algorithm with small *expected* running time that is *guaranteed* to terminate in poly-time. This is achieved by just running the probabilistic algorithm in parallel to the algorithm from Theorem 1 and terminating after at least one of the two succeeds.

**Corollary 2** *Suppose that there is a probabilistic ad hoc routing algorithm that achieves an expected routing time of $T(n)$ such that its failure probability is $n^{-\omega(1)}$ (that is smaller than inverse polynomial), then there is a probabilistic ad hoc routing algorithm that achieves and expected routing time of $O(T(n))$ and is* guaranteed *to succeed if a path exists.*

A similar statement can be made about broadcasting. In addition, since the algorithm works for *all* graphs, a similar statement can be made where we are given a routing algorithm that works well for most graphs drawn according to some distribution.

## 2 Universal exploration sequences

Suppose that $G = (V, E)$ is an undirected graph with degree bounded by $D$. Further, at each vertex $v$ each of the edges $(v, u)$ leaving $v$ are assigned labels $l(v, u)$ which are a permutation of $0, 1, \ldots, deg(v) - 1$ in an arbitrary way. Thus the "exits" from $v$ are assigned unique labels. The labels of an edge $(u, v)$ from the viewpoint of $u$ and $v$ do not necessarily have to match.

An exploration sequence is a set of integer "directions" $t_1, t_2, \ldots$ on how to traverse the graph. If before step $i$ the walk enters a vertex $v$ on edge $(u, v)$ then it would leave on edge $(v, w)$ such that $l(v, w) = l(v, u) + t_i \bmod d(v)$ (cf. [6]). Note that the definition is meaningful even when the graph $G$ is not regular. By applying degree reduction as in [6], we will only need the notion for 3-regular graphs. An important property of exploration sequences is that they are *reversible*. That is, if one knows $t_i$ and that on step $i$ the walk took the edge $(v, w)$ then one can infer that on step $i - 1$ the walk took the edge $(u, v)$ such that $l(v, u) = l(v, w) - t_i \bmod d(v)$.

From now on we will focus on 3-regular graphs. Given a 3-regular labeled graph $G$, a starting edge $e_0$ and an exploration sequence $t_1, t_2, \ldots \in \{0, 1, 2\}$, we can trace the walk $e_0, e_1, e_2, \ldots$ dictated by this sequence. One hopes that if the sequence is long enough all the vertexes would eventually be visited, irrespective of the labeling or the starting vertex. *Universal exploration sequences* are objects that guarantee this property.

**Definition 3** *A sequence $T_n = \{t_1, \ldots, t_N\}$ is a universal exploration sequence for connected 3-regular graphs of size $\leq n$ if for any connected 3-regular graph $G = (V, E)$ with $|V| \leq n$, for any labeling and for any initial edge $e \in E$, following the sequence as dictated by $T$ will result in visiting all vertexes in the graph.*

It can be shown that a sufficiently long random walk – one of length $O(n^2)$ in our case [3, 7] will cover any given 3-regular $G$ with high probability. Via a probabilistic argument, one can show that almost *any* sufficiently long sequence $T$ in $\{0, 1, 2\}^*$ is a universal exploration sequence. Moreover $N = |T|$ can be chosen $poly(n)$. It is a much trickier question whether a *specific* universal exploration sequence can be constructed. Recently, Reingold [8] demonstrated that this is the case and, moreover, these sequences can be constructed in logspace.

**Theorem 4** *([8], Cor. 5.5) There exists a log-space algorithm that takes as input $1^n$ and produces a universal exploration sequence $T_n$ for 3-regular graphs of size $\leq n$.*

Denote the length of the sequence from Theorem 4 by $L_n = |T_n|$. Since the algorithm is log-space it follows that $L_n < poly(n)$. Note that there is no need to store the universal exploration sequence (which we wouldn't be able to do in log-space), since for any $1 \leq j \leq L_n$ the algorithm from Theorem 4 would be able to compute the $j$-th element of the sequence in log-space, by repeating the computation and ignoring the output until the $j$-th element $T_n[j]$ of the output is reached. Assuming that the walk arrived at vertex $v$ from vertex $u$, we will be able to compute the next vertex

$$w = next_v((u,v), T_n[j]) := w \text{ such that } l(v,w) = l(v,u) + T_n[j] \ (mod \ d(v)).$$

Similarly, by the reversibility of the universal exploration sequence, knowing $w$ we will be able to compute

$$u = prev_v((v,w), T_n[j]) := u \text{ such that } l(v,u) = l(v,w) - T_n[j] \ (mod \ d(v)).$$

Both computations can obviously be done in space $O(\log n)$.

## 3 Guaranteed routing

A message has to be delivered from node $s$ to node $t$ in the graph. We would like the message to be delivered in a decentralized fashion, in a way that would guarantee delivery. A natural approach would be to use the universal exploration sequence originated from $s$. It is guaranteed to reach $t$ at some point. We would like the origin node $s$ to learn that transmission has failed if the message has not reached $t$.

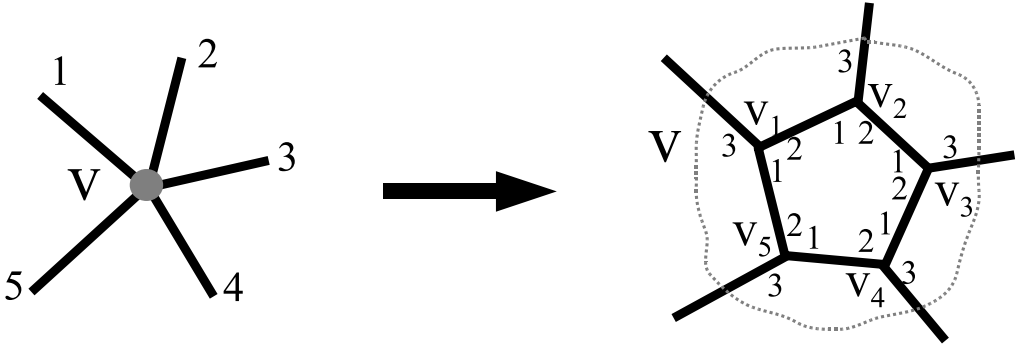

Figure 1: Reducing the degree to 3

A first technical step would be to perform a degree reduction to convert the graph $G$ into a 3-regular graph $G'$. The conversion is fairly standard (e.g. see [6], p. 80), and is illustrated on Fig. 1. Each node $v$ "simulates" $O(deg(v))$ nodes of degree 3, thus at most squaring the size of the graph. We first assume that an upper bound $n$ on the number of nodes in the connected component of $s$ in $G'$ is known. The routing will run in time $poly(n)$ and each node will use $O(\log n)$ space. Later we will show how to drop the assumption that $n$ is known ahead of time. Recall that the universal exploration sequence guaranteed by Theorem 4 is denoted by $T_n$ and $L_n = |T_n| = poly(n)$.

The message's header will be of size $O(\log n)$ and will contain the following information: the name of the source node $s$, the name of the destination node $t$, direction $dir$ (one bit, will be used to send confirmation back to $s$), $status$ (one bit, success or failure), and the index $i \in [1..L_n]$ which defines where along the universal exploration path the message currently is. The algorithm works as follows:

> **Algorithm** *Route*
> when node $v$ receives message $m$ from node $u$ with header
> $(s, t, dir, status, i)$ do the following:
>  **if** $dir =$ "*back*" and $v = s$
>   **return** *status*;
>  **if** $dir =$ "*forward*" and $v = t$
>   $dir :=$ "*back*", $i := i - 1$, $status :=$ *success*, send message back to $u$;
>  **if** $d =$ "*forward*" and $i > L_n$
>   $dir :=$ "*back*", $i := i - 1$, $status :=$ *failure*, send message back to $u$;
>  compute $T_n[i]$;
>  **if** $d =$ "*forward*"
>   $i := i + 1$, send $m$ to $next_v((u, v), T_n[i])$;
>  **else**
>   $i := i - 1$, send $m$ to $prev_v((v, u), T_n[i])$;

Assuming the connected component of $s$ in $G'$ has at most $n$ vertexes, the universal exploration sequence is guaranteed to eventually visit them all. If $t$ is in the connected component, then it will be encountered, and after this happens, the algorithm will send a confirmation back to $s$ and return "*success*". If after $L_n$ steps $t$ sill hasn't been encountered, it must not be connected to $s$ and the algorithm will return "*failure*", after backtracking back to $s$ along $T_n$. Each node uses logarithmic space to compute $T_n[i]$, and the number of steps in the route is bounded by $poly(n)$.

## 4 Counting the number of vertexes

In this section we show that the number $n$ of vertexes in the connected component of $s$ in $G'$ can be computed in time $poly(n)$, thus showing that guaranteed ad hoc routing can be achieved in polynomial time without any prior knowledge on the network (not even the size). The idea is as follows, we run universal exploration sequences from $s$ of $T_1, T_2, T_4, \ldots$ and see whether the walk covers the entire connected component of $s$, that is whether all the neighbors of vertexes visited by $T_{2^k}$ are also visited by $T_{2^k}$. Note that running $T_{2^k}$ takes time $poly(2^k)$, and if the size of the connected component $\mathcal{C}_s$ of $s$ is $n$, the walk $T_{2^k}$ for some $2^k < 2n$ will visit the entire component. Thus the counting is done in time polynomial in $n = |\mathcal{C}_s|$.

We first show how to check whether $T_{2^k}$ originating at $s$ covers $\mathcal{C}_s$. For any $i \leq L_{2^k}$, $s$

can receive the identifier for $v_i$, the node visited on the $i$-th step when $T_{2^k}$ is followed from $s$. We denote this response by $Retrieve(s, T_{2^k}, i)$. This command only requires indexes ($O(k)$ space) and the identifier of one vertex ($v_i$) to be transmitted. Similarly, we can retrieve the $j$-th neighbor $u_i^j$ of $v_i$ ($j = 1, 2, 3$), using the same amount of memory. We denote the response by $RetrieveNeighbor(s, T_{2^k}, i, j)$. We can then compare $u_i^j$ to $v_\ell$ for $\ell = 1, \ldots, L_{2^k}$ to see whether $u_i^j$ is already visited by the walk.

---

**Algorithm** $CountNodes(s)$
$k = 0$;
**do**
    $k := k + 1$, $NewNodeDiscovered := false$;
    compute $L_{2^k} = |T_{2^k}|$;
    **for** $i := 1$ **to** $L_{2^k}$ **do**
        **for** $j := 1$ **to** $3$ **do**
            $u_i^j := RetrieveNeighbor(s, T_{2^k}, i, j)$;
            $uIsNew := true$;
            **for** $\ell := 1$ **to** $L_{2^k}$ **do**
                $v_\ell := Retrieve(s, T_{2^k}, \ell)$;
                **if** $v_\ell = u_i^j$
                    $uIsNew := false$;
            **if** $uIsNew = true$
                $NewNodeDiscovered := true$, skip to "while";
**while** $NewNodeDiscovered = true$;
$NodeCount := 0$;
**for** $i := 1$ **to** $L_{2^k}$ **do**
    $newnode := \text{"}true\text{"}$;
    $v_{new} := Retrieve(s, T_{2^k}, i)$;
    **for** $j := 1$ **to** $i - 1$ **do**
        $v_{old} := Retrieve(s, T_{2^k}, j)$;
        **if** $v_{old} = v_{new}$
            $newnode := \text{"}false\text{"}$;
    **if** $newnode = \text{"}true\text{"}$
        $NodeCount := NodeCount + 1$;
**return** $NodeCount$;

---

Denote the nodes visited by $T_{2^k}$ by $V_k$. If all the nodes adjacent to those in $V_k$ are also in $V_k$, it means that $V_k$ form a connected component and thus $V_k = \mathcal{C}_s$. The algorithm thus first makes sure that $T_{2^k}$ visits all the nodes in $\mathcal{C}_s$, and then counts the number of nodes it visits. The algorithm runs in time $O(L_{2^k}^2) = poly(2^k) = poly(|\mathcal{C}_s|)$, and space logarithmic as described above.

## Acknowledgments

The author wishes to thank Maia Fraser and Stephane Durocher for the insightful conversations during the preparation of this work.